\documentclass[12pt]{article}
\usepackage{graphicx}
\usepackage{color}
\def\hybrid{\topmargin 0pt      \oddsidemargin 0pt
        \headheight 0pt \headsep 0pt
       \voffset-1cm
        \textwidth 6.25in       % A4 paper
       \textheight 9.5in       % A4 paper
        \marginparwidth 0.0in
        \parskip 5pt plus 1pt   \jot = 1.5ex}
\catcode`\@=11
\def\marginnote#1{}

\newcount\hour
\newcount\minute
\newtoks\amorpm
\hour=\time\divide\hour by60
\minute=\time{\multiply\hour by60 \global\advance\minute by-\hour}
\edef\standardtime{{\ifnum\hour<12 \global\amorpm={am}%
        \else\global\amorpm={pm}\advance\hour by-12 \fi
        \ifnum\hour=0 \hour=12 \fi
        \number\hour:\ifnum\minute<10 0\fi\number\minute\the\amorpm}}
\edef\militarytime{\number\hour:\ifnum\minute<10 0\fi\number\minute}

\def\draftlabel#1{{\@bsphack\if@filesw {\let\thepage\relax
   \xdef\@gtempa{\write\@auxout{\string
      \newlabel{#1}{{\@currentlabel}{\thepage}}}}}\@gtempa
   \if@nobreak \ifvmode\nobreak\fi\fi\fi\@esphack}
        \gdef\@eqnlabel{#1}}
\def\@eqnlabel{}
\def\@vacuum{}
\def\draftmarginnote#1{\marginpar{\raggedright\scriptsize\tt#1}}

\def\draftlabel#1{{\@bsphack\if@filesw {\let\thepage\relax
   \xdef\@gtempa{\write\@auxout{\string
      \newlabel{#1}{{\@currentlabel}{\thepage}}}}}\@gtempa
   \if@nobreak \ifvmode\nobreak\fi\fi\fi\@esphack}
        \gdef\@eqnlabel{#1}}
\def\@eqnlabel{}
\def\@vacuum{}
\def\draftmarginnote#1{\marginpar{\raggedright\scriptsize\tt#1}}

\def\draft{\oddsidemargin -.5truein
        \def\@oddfoot{\sl preliminary draft \hfil
        \rm\thepage\hfil\sl\today\quad\militarytime}
        \let\@evenfoot\@oddfoot \overfullrule 3pt
        \let\label=\draftlabel
        \let\marginnote=\draftmarginnote
   \def\@eqnnum{(\theequation)\rlap{\kern\marginparsep\tt\@eqnlabel}%
\global\let\@eqnlabel\@vacuum}  }

\def\numberbysection{\@addtoreset{equation}{section}
        \def\theequation{\thesection.\arabic{equation}}}

\def\underline#1{\relax\ifmmode\@@underline#1\else
        $\@@underline{\hbox{#1}}$\relax\fi}

\def\titlepage{\@restonecolfalse\if@twocolumn\@restonecoltrue\onecolumn
     \else \newpage \fi \thispagestyle{empty}\c@page\z@
        \def\thefootnote{\fnsymbol{footnote}} }

\def\endtitlepage{\if@restonecol\twocolumn \else  \fi
        \def\thefootnote{\arabic{footnote}}
        \setcounter{footnote}{0}}  %\c@footnote\z@ }
\relax

\hybrid

\newfont{\Bbb}{msbm10 scaled 1\@ptsize00}
\newfont{\Bbbb}{msbm7 scaled 1\@ptsize00}

\newcommand{\DDD}{\raise-1pt\hbox{$\mbox{\Bbbb D}$}}

\newcommand{\UUU}{\raise-1pt\hbox{$\mbox{\Bbbb U}$}}

\newcommand{\ZZ}{\mbox{\Bbb Z}}
\newcommand{\z}{\raise-1pt\hbox{$\mbox{\Bbbb Z}$}}

\def\beq{\begin{equation}}
\def\eeq{\end{equation}}
\def\p{\partial}

\def\normord{ {\scriptstyle {{\bullet}\atop{\bullet}}} }
\def\lbr{\left <}
\def\rbr{\right >}

\begin{document}

\begin{titlepage}

\title{On matrix modified KP hierarchy}

\author{A.~Zabrodin\thanks{National Research University Higher School of Economics,
20 Myasnitskaya Ulitsa, Moscow 101000, Russian Federation;
ITEP, 25 B.Cheremushkinskaya, Moscow 117218, Russian Federation;
Skolkovo Institute of Science and Technology, 143026 Moscow, Russian Federation;
e-mail: zabrodin@itep.ru}
}

\date{February 2018}
\maketitle

\vspace{-7cm} \centerline{ \hfill ITEP-TH-04/18}\vspace{7cm}

\begin{abstract}

Using the bilinear formalism, 
we consider multicomponent and matrix modified KP hierarchies.
The main tool is the bilinear identity
for the tau-function which is realized as an expectation value of 
a Clifford group element composed from multicomponent fermionic operators. 
We also construct the Baker-Akhiezer functions and derive auxiliary linear
equations satisfied by them. 

\end{abstract}

\end{titlepage}

\vspace{5mm}

\section{Introduction}

Integrable hierarchies of non-linear partial differential and difference
equations are in the basis of the theory of integrable systems.
Non-linear equations of integrable hierarchies can be represented as compatibility
conditions for certain auxiliary linear problems. Special solutions to these linear
problems depending on a complex spectral parameter 
$z$ are called Baker-Akhiezer functions $\Psi =\Psi (z)$. 
The multicomponent integrable hierarchies (see 
\cite{DJKM81,KL93,Teo11,TT07}) are usually formulated in a matrix form
with matrix pseudodifferential operators and matrix-valued Baker-Akhiezer functions.

Among different known examples of integrable hierarchies an archetypal one is the
Kadomtsev-Petviashvili (KP) hierarchy.  
The modified KP (mKP) hierarchy is a larger hierarchy than the KP one.
The set of independent variables in the $N$-component mKP hierarchy
consists in $N$ infinite sets of continuous time variables $t_{\alpha , m}$
($\alpha =1, \ldots , N$, $m=1, 2, \ldots $) and a finite set of $N$ auxiliary discrete
variables $p_1, \ldots , p_N$ ($p_{\alpha}\in \ZZ$). The restriction to the 
(multicomponent) KP
hierarchy is achieved by fixing the $p$-variables to zero values.
What is usually called matrix mKP hierarchy is a restriction of the 
multicomponent mKP hierarchy to the following values of the times:
$t_{\alpha , m}=t_m$ for each $\alpha$ and $m$,
$p_{\alpha}=p$ for each $\alpha$.

In this paper we are going to discuss some aspects of the theory of the 
multicomponent and matrix mKP hierarchies which seem to be missing in the literature.
We introduce the wave operator and 
derive the auxiliary linear problems in the form
$\p_{t_{\alpha , m}}\Psi = A_{\alpha m}\Psi$, where $A_{\alpha m}$ are matrix difference 
operators in $p$ of order $m$ (linear combinations of the shift operators 
$e^{k\p_p}$ with $k=0, \ldots , m$ with coefficients depending on the times).
For the matrix mKP hierarchy, the linear problems become 
$\p_{t_m}\Psi = A_{m}\Psi$. In particular, we obtain from the first principles 
the linear problem $\p_{t_1}\Psi =e^{\p_p}\Psi +u\Psi$, where $u$ is a certain matrix
function. We also derive the auxiliary linear problems for the adjoint Baker-Akhiezer
function $\Psi^{\dag}$. 
What is more, we obtain the Lax representation of the matrix mKP hierarchy, in which 
the Lax operator is a pseudo-difference operator in $p$, i.e., an infinite linear combination
of $e^{k\p_p}$ with $k\leq 1$ with coefficients depending on the times.

Our starting point is the formalism of free fermions developed in 
\cite{DJKM83,JM83}. For the $N$-component hierarchy one needs $N$-component
fermionic fields $\psi^{(\alpha )}(z)$, $\psi^{*(\alpha )}(z)$. The main object
of the theory is the tau-function $\tau$ which is a vacuum expectation value of certain
composite fermion operators $g$ belonging to the Clifford group.
In the $N$-component case, it is natural to consider an array of tau-functions
$\tau_{\alpha \beta}$ which are
matrix elements of operators from the Clifford group between different vacua
with the same total charge.
The key identity that allows one to derive equations of the hierarchy is the
bilinear identity in the operator form 
$$
\sum_{\gamma =1}^N \oint \frac{dz}{z} \, \psi^{(\gamma )}(z)g\otimes \psi^{*(\gamma )}(z)g=
\sum_{\gamma =1}^N \oint \frac{dz}{z} \, g\psi^{(\gamma )}(z)\otimes g\psi^{*(\gamma )}(z),
$$
which holds for any $g$ from the Clifford group. Taking matrix elements
of the operator bilinear identity between appropriate states, one is able to derive 
the fundamental bilinear identity for the tau-functions which is the corner stone
of our approach. The bilinear identity generates a number of bilinear
equations for the tau-functions of the Hirota type. 

The plan of the paper is as follows. In section 2 we introduce the multicomponent
fermions $\psi^{(\alpha )}(z)$, $\psi^{*(\alpha )}(z)$, 
modes of the current operator $J_{m}^{(\alpha )}$ 
coupled with the time variales $t_{\alpha , m}$
and define the tau-function as an expectation value of a Clifford group element.
Using the bosonization rules, the bilinear identity for the array of tau-functions
of the multicomponent mKP hierarchy is derived and some of its important
consequences are explicitly written. Next, we introduce the Baker-Akhiezer
function $\Psi$ and its adjoint $\Psi^{\dag}$ 
in terms of the tau-functions. In section 3 we
consider the specialization to the matrix mKP hierarchy. We introduce the 
pseudo-difference wave operator and derive the auxiliary linear problems 
for the Baker-Akhiezer function and its adjoint. Section 4 is devoted to 
the special linear problem for the time $t_1$. By a direct calculation, we
show that it follows from the bilinear identity for the tau-functions.

\section{Multicomponent fermions and bilinear identity}

\subsection{The multicomponent fermions}

Following \cite{DJKM81,TT07}, 
we introduce the creation-annihilation multicomponent free fermionic operators 
labeled by $\alpha =1, \ldots , N$ as $\psi_{j}^{(\alpha )}$,
$\psi_{j}^{*(\alpha )}$ ($j\in \ZZ$). They obey the anti-commutation relations
$$
[\psi_{j}^{(\alpha )}, \psi_{k}^{*(\beta )}]_+=\delta_{\alpha \beta}\delta_{jk},
\qquad
[\psi_{j}^{(\alpha )}, \psi_{k}^{(\beta )}]_+=
[\psi_{j}^{*(\alpha )}, \psi_{k}^{*(\beta )}]_+=0.
$$
The Fock and dual Fock spaces are generated by the vacuum states 
$\left | {\bf 0}\rbr$, $\lbr {\bf 0} \right |$ that satisfy the conditions
$$
\psi_{j}^{(\alpha )}\left | {\bf 0}\rbr =0 \quad (j<0), \qquad
\psi_{j}^{*(\alpha )}\left | {\bf 0}\rbr =0 \quad (j\geq 0),
$$
$$
\lbr {\bf 0}\right | \psi_{j}^{(\alpha )} =0 \quad (j\geq 0), \qquad
\lbr {\bf 0}\right | \psi_{j}^{*(\alpha )} =0 \quad (j< 0),
$$
so $\psi_{j}^{(\alpha )}$ with $j<0$ and $\psi_{j}^{*(\alpha )}$ with
$j\geq 0$ are annihilation operators while 
$\psi_{j}^{(\alpha )}$ with $j\geq 0$ and
$\psi_{j}^{*(\alpha )}$ with
$j<0$ are creation operators. Let ${\bf p}=(p_1, p_2, \ldots , p_N)$ be a set
of integer numbers. We define the states $\left | {\bf p}\rbr$, $\lbr {\bf p} \right |$
as
$$
\left | {\bf p}\rbr =\Psi_{p_N}^{*(N)}\ldots \Psi_{p_2}^{*(2)}
\Psi_{p_1}^{*(1)}\left | {\bf 0}\rbr , \qquad
\lbr {\bf p} \right |=\lbr {\bf 0} \right |\Psi_{p_1}^{(1)}\Psi_{p_2}^{(2)}\ldots
\Psi_{p_N}^{(N)},
$$
where
$$
\Psi_{p}^{*(\alpha )}=\left \{ \begin{array}{l}
\psi^{(\alpha )}_{p-1}\ldots \psi^{(\alpha )}_{0} \quad \,\,\, (p >0)
\\
\psi^{*(\alpha )}_{p}\ldots \psi^{*(\alpha )}_{-1} \quad (p <0),
\end{array}
\right.
$$
$$
\Psi_{p}^{(\alpha )}=\left \{ \begin{array}{l}
\psi^{*(\alpha )}_{0}\ldots \psi^{*(\alpha )}_{p-1} \quad  \, (p >0)
\\
\psi^{(\alpha )}_{-1}\ldots \psi^{(\alpha )}_{p} \quad \,\,\,\,\, (p <0).
\end{array}
\right.
$$

Let us introduce the operators
$$
J_{k}^{(\alpha )}=\sum_{j\in \z}\normord \psi^{(\alpha )}_{j} \psi^{*(\alpha )}_{j+k}
\normord ,
$$
where the normal ordering is defined by moving the annihilation operators 
to the right and creation operators to the left with the minus sign emerging each time
when two fermionic operators are permuted (in fact the normal ordering
is essential only for $J_0^{(\alpha )}$). They are Fourier modes of the current
operator. The operators $J_{0}^{(\alpha )}=Q_{\alpha}$ are charge operators. 
Assuming that there are $N$ 
infinite sets of the independent continuous time
variables
$$
{\bf t}=\{{\bf t}_1, {\bf t}_2, \ldots , {\bf t}_N\}, \qquad
{\bf t}_{\alpha}=\{t_{\alpha , 1}, t_{\alpha , 2}, t_{\alpha , 3}, \ldots \, \},
\qquad \alpha = 1, \ldots , N,
$$
we introduce the operator
$$
J({\bf t})=\sum_{\alpha =1}^N \sum_{k\geq 1} t_{\alpha , k}J_k^{(\alpha )}.
$$

The tau-function $\tau ({\bf p}, {\bf t})$ of the multicomponent  
mKP hierarchy is defined as the expectation value
\beq\label{f1}
\tau ({\bf p}, {\bf t})=\lbr {\bf p}\right | e^{J({\bf t})} g\left |{\bf p}\rbr ,
\eeq
where $g$ is a general element of the Clifford group whose typical form is
$$
g=\exp \left ( \sum_{\alpha , \beta}\sum_{j,k}A_{jk}^{(\alpha \beta )}
\psi^{(\alpha )}_{j}\psi^{*(\beta )}_{k}\right )
$$
with some infinite matrix $A_{jk}^{(\alpha \beta )}$.

\subsection{The bilinear identity}

An important property of the Clifford group elements is the following 
operator bilinear identity:
\beq\label{f2}
\sum_{\gamma =1}^N \sum_{j\in \z}\psi_{j}^{(\gamma )}g \otimes
\psi_{j}^{*(\gamma )}g =\sum_{\gamma =1}^N \sum_{j\in \z}
g\psi_{j}^{(\gamma )}\otimes g \psi_{j}^{(\gamma )}.
\eeq
Let us introduce the free fermionic fields
$$
\psi^{(\alpha )}(z)=\sum_{j\in \z}\psi^{(\alpha )}_j z^j,
\qquad
\psi^{*(\alpha )}(z)=\sum_{j\in \z}\psi^{*(\alpha )}_j z^{-j},
$$
then the operator bilinear identity acquires the form
\beq\label{f3}
\sum_{\gamma =1}^N \oint \frac{dz}{z} \, \psi^{(\gamma )}(z)g\otimes \psi^{*(\gamma )}(z)g=
\sum_{\gamma =1}^N \oint \frac{dz}{z} \, g\psi^{(\gamma )}(z)\otimes g\psi^{*(\gamma )}(z).
\eeq
Here the contour integral is understood to be an integral along the big circle
$|z|=R$ with sufficiently large $R$, $\oint dz z^n =2\pi i \delta_{n, -1}$.

The key identity is
$$
\psi_{j}^{(\gamma )}\left |{\bf p}\rbr \otimes \psi_{j}^{*(\gamma )}\left |{\bf p}'\rbr =0
\qquad \mbox{if $p_{\alpha}\geq p_{\alpha}'$ for all $\alpha$},
$$
which holds for all $j\in \ZZ$ because either $\psi_{j}^{(\gamma )}$ annihilates
$\left |{\bf p}\rbr$ or
$\psi_{j}^{*(\gamma )}$ annihilates $\left |{\bf p}'\rbr$. Therefore, applying
both sides of (\ref{f2}) to $\left |{\bf p}\rbr \otimes \left |{\bf p}'\rbr$, we get
$$
\sum_{\gamma =1}^N \sum_{j\in \z}\psi_{j}^{(\gamma )}g \left |{\bf p}\rbr \otimes
\psi_{j}^{*(\gamma )}g \left |{\bf p}'\rbr =0
\qquad \mbox{if $p_{\alpha}\geq p_{\alpha}'$ for all $\alpha$}
$$
or
\beq\label{f4}
\sum_{\gamma =1}^N \oint \frac{dz}{z} \, 
\psi^{(\gamma )}(z)g\left |{\bf p}\rbr \otimes \psi^{*(\gamma )}(z)g\left |{\bf p}'\rbr =0
\qquad \mbox{if $p_{\alpha}\geq p_{\alpha}'$ for all $\alpha$}.
\eeq
Now apply $\lbr {\bf p}+{\bf e}_{\alpha}\right |e^{J({\bf t})}\otimes 
\lbr {\bf p}'-{\bf e}_{\beta}\right |e^{J({\bf t}')}$, where 
${\bf e}_{\alpha}$ denotes the vector with 1 on the $\alpha$th place and zeros
elsewhere, to get
$$
\sum_{\gamma =1}^N \oint \frac{dz}{z}  
\lbr {\bf p}\! +\! {\bf e}_{\alpha}\right |e^{J({\bf t})}
\psi^{(\gamma )}(z)g\left |{\bf p}\rbr 
\lbr {\bf p}'\! -\! {\bf e}_{\beta}\right |e^{J({\bf t}')}
\psi^{*(\gamma )}(z)g\left |{\bf p}'\rbr =0
\quad \mbox{if $p_{\alpha}\geq p_{\alpha}'$ for all $\alpha$}.
$$
or
\beq\label{f6}
\sum_{\gamma =1}^N \oint \frac{dz}{z} \, e^{\xi ({\bf t}_{\gamma}-{\bf t}_{\gamma}', z)} 
\lbr {\bf p}\! +\! {\bf e}_{\alpha} \right | \psi^{(\gamma )}(z)e^{J({\bf t})}
g\left |{\bf p}\rbr 
\lbr {\bf p}'\! -\! {\bf e}_{\beta} \right | \psi^{*(\gamma )}(z) e^{J({\bf t}')}
g\left |{\bf p}'\rbr =0.
\eeq
Here
\beq\label{f5}
\xi ({\bf t}_{\gamma}, z)=\sum_{k\geq 1}t_{\gamma , k}z^k
\eeq
and the commutation relations
$$
e^{J({\bf t})}\psi^{(\gamma )}(z)=e^{\xi ({\bf t}_{\gamma}, z)}
\psi^{(\gamma )}(z)e^{J({\bf t})}, \qquad
e^{J({\bf t})}\psi^{*(\gamma )}(z)=e^{-\xi ({\bf t}_{\gamma}, z)}
\psi^{*(\gamma )}(z)e^{J({\bf t})}
$$
are used. Now we are ready to employ the multicomponent bosonization rules \cite{KL93}
$$
\lbr {\bf p}\! +\! {\bf e}_{\alpha} \right | \psi^{(\gamma )}(z)e^{J({\bf t})}=
\epsilon_{\alpha \gamma}\epsilon_{\gamma}({\bf p})z^{p_{\gamma}+\delta_{\alpha \gamma} -1}
\lbr {\bf p}\! +\! {\bf e}_{\alpha}\! -\! {\bf e}_{\gamma}\right |
e^{J({\bf t}-[z^{-1}]_{\gamma})},
$$
$$
\lbr {\bf p}'\! -\! {\bf e}_{\beta} \right | \psi^{*(\gamma )}(z)e^{J({\bf t})}=
\epsilon_{\beta \gamma}\epsilon_{\gamma}({\bf p}')z^{p_{\gamma}'+\delta_{\beta \gamma}}
\lbr {\bf p}'\! -\! {\bf e}_{\beta}\! +\! {\bf e}_{\gamma}\right |
e^{J({\bf t}+[z^{-1}]_{\gamma})},
$$
where
$$
\left ({\bf t}\pm [z^{-1}]_{\gamma}\right )_{\alpha k}=t_{\alpha , k}\pm
\delta_{\alpha \gamma} \frac{z^{-k}}{k}
$$
and the sign factors $\epsilon_{\alpha \beta}$, $\epsilon_{\gamma}({\bf p})$ are:
$\epsilon_{\alpha \beta}=1$ if $\alpha \leq \beta$, 
$\epsilon_{\alpha \beta}=-1$ if $\alpha > \beta$,
$\epsilon_{\gamma}({\bf p})=(-1)^{p_{\gamma +1}+\ldots + p_N}$.
Multiplying (\ref{f6}) by $\epsilon_{\alpha}({\bf p})\epsilon_{\beta}({\bf p}')$,
we arrive at the bilinear identity for the tau-function of the $N$-component
mKP hierarchy
\beq\label{f7}
\sum_{\gamma =1}^N \epsilon_{\alpha \gamma}({\bf p})\epsilon_{\beta \gamma}({\bf p}')
\! \oint dz z^{p_{\gamma}-p'_{\gamma}+\delta_{\alpha \gamma}+\delta_{\beta \gamma}-2}
e^{\xi ({\bf t}_{\gamma}-{\bf t}_{\gamma}', z)} 
\tau_{\alpha \gamma}({\bf p}, {\bf t}-[z^{-1}]_{\gamma})
\tau_{\gamma \beta}({\bf p}', {\bf t}'+[z^{-1}]_{\gamma})=0
\eeq
valid for any ${\bf t}$, ${\bf t}'$, 
${\bf p}$, ${\bf p}'$ if $p_{\alpha}\geq p_{\alpha}'$ for all $\alpha$. Here
and below
$$
\tau_{\alpha \gamma}({\bf p}, {\bf t})=
\lbr {\bf p}\! +\! {\bf e}_{\alpha}\! -\! {\bf e}_{\gamma}\right |
e^{J({\bf t})}g\left |{\bf p}\rbr
$$
and
$$
\epsilon_{\alpha \gamma}({\bf p})=\left \{
\begin{array}{ll}
\;\; (-1)^{p_{\alpha +1}+\ldots +p_{\gamma}} &\quad \mbox{if $\alpha <\gamma$}
\\
\quad 1 &\quad \mbox{if $\alpha =\gamma$}
\\
-(-1)^{p_{\gamma +1}+\ldots +p_{\alpha}} &\quad \mbox{if $\alpha >\gamma$}
\end{array}\right.
$$
The integration contour around $\infty$ is such that all singularities 
coming from the power of $z$ and the exponential function
$e^{\xi ({\bf t}_{\gamma}-{\bf t}_{\gamma}', \, z)}$ are inside it and all singularities 
coming from the $\tau$-factors are outside it. 

\subsection{The Hirota equations}

At ${\bf p}={\bf p}'$ the bilinear identity generates the
$N$-component KP hierarchy. Choosing ${\bf t}'$ in (\ref{f7}) in a specific way, 
one can obtain, after calculating the residues, a number of differential and difference 
bilinear equations for the tau-function of the Hirota type (they are called Fay
identities in \cite{Teo11}). The complete list of such equations is given
in \cite{Teo11}. Below we present only the equations that are used in what follows.

Differentiating (\ref{f7}) (with ${\bf p}={\bf p}'$) 
with respect to $t_{\gamma , 1}$ and setting
${\bf t}'={\bf t}-[\mu^{-1}]_{\beta}$, we have, for any distinct $\alpha , \beta , \gamma$:
\beq\label{f8}
\tau_{\alpha \beta}({\bf t}\! -\! [\mu^{-1}]_{\beta})\p_{t_{\gamma , 1}}\tau ({\bf t})-
\tau ({\bf t})\p_{t_{\gamma , 1}}\tau_{\alpha \beta}({\bf t}\! -\! [\mu^{-1}]_{\beta})
+\frac{\epsilon_{\alpha \gamma}({\bf p})
\epsilon_{\gamma \beta}({\bf p})}{\epsilon_{\alpha \beta}({\bf p})}\,
\tau_{\alpha \gamma}({\bf t})\tau_{\gamma \beta}({\bf t}\! -\! [\mu^{-1}]_{\beta})=0,
\eeq
where we have suppressed the dependence of the tau-function on ${\bf p}$ (since
${\bf p}$ is the same for all tau-functions).  

Differentiating (\ref{f7}) with respect to $t_{\beta , 1}$ and setting
${\bf t}'={\bf t}-[\mu^{-1}]_{\alpha}-[\nu^{-1}]_{\beta}$,
we have, for any distinct $\alpha , \beta$:
\beq\label{f9}
\begin{array}{c}
\p_{t_{\beta , 1}}\tau_{\alpha \beta}({\bf t}-[\nu^{-1}]_{\beta})
\tau ({\bf t}-[\mu^{-1}]_{\alpha})-
\p_{t_{\beta , 1}}\tau ({\bf t}-[\mu^{-1}]_{\alpha})
\tau_{\alpha \beta}({\bf t}-[\nu^{-1}]_{\beta})
\\ \\
+\nu \tau_{\alpha \beta}({\bf t}-[\nu^{-1}]_{\beta})
\tau ({\bf t}-[\mu^{-1}]_{\alpha})
-\nu \tau_{\alpha \beta}({\bf t})\tau ({\bf t}-[\mu^{-1}]_{\alpha}-[\nu^{-1}]_{\beta})=0.
\end{array}
\eeq
In a similar way, differentiating (\ref{f7}) with respect to $t_{\alpha , 1}$ and setting
${\bf t}'={\bf t}-[\mu^{-1}]_{\alpha}-[\nu^{-1}]_{\beta}$,
we have, for any distinct $\alpha , \beta$:
\beq\label{f10}
\begin{array}{c}
\p_{t_{\alpha , 1}}\tau_{\alpha \beta}({\bf t}-[\nu^{-1}]_{\beta})
\tau ({\bf t}-[\mu^{-1}]_{\alpha})-
\p_{t_{\alpha , 1}}\tau ({\bf t}-[\mu^{-1}]_{\alpha})
\tau_{\alpha \beta}({\bf t}-[\nu^{-1}]_{\beta})
\\ \\
-\mu \tau_{\alpha \beta}({\bf t}-[\nu^{-1}]_{\beta})
\tau ({\bf t}-[\mu^{-1}]_{\alpha})
+\mu \tau ({\bf t})\tau_{\alpha \beta}({\bf t}-[\mu^{-1}]_{\alpha}-[\nu^{-1}]_{\beta})=0.
\end{array}
\eeq

Differentiating (\ref{f7}) at $\beta =\alpha$ 
with respect to $t_{\gamma , 1}$ ($\gamma \neq \alpha$) and setting
${\bf t}'={\bf t}-[\mu^{-1}]_{\alpha}$,
we have, for any distinct $\alpha , \gamma$:
\beq\label{f11}
\p_{t_{\gamma , 1}}\tau ({\bf t}-[\mu^{-1}]_{\alpha})\tau ({\bf t})-
\p_{t_{\gamma , 1}}\tau ({\bf t})\tau ({\bf t}-[\mu^{-1}]_{\alpha})+
\mu^{-1}\tau_{\alpha \gamma}({\bf t})\tau_{\gamma \alpha}({\bf t}-[\mu^{-1}]_{\alpha})=0.
\eeq

Consequences of the bilinear identity with ${\bf p}\neq {\bf p}'$ will be discussed
in the next sections.

\subsection{The Baker-Akhiezer functions}

The matrix Baker-Akhiezer function $\Psi ({\bf p}, {\bf t};z)$ and its adjoint
$\Psi ^{\dag}({\bf p}, {\bf t};z)$ are $N\! \times \! N$ matrices with components
defined by 
\beq\label{b1}
\begin{array}{l}
\displaystyle{\Psi_{\alpha \beta}({\bf p}, {\bf t};z)=
\epsilon_{\alpha \beta}({\bf p})\,
\frac{\tau_{\alpha \beta}({\bf p}, {\bf t}-[z^{-1}]_{\beta})}{\tau ({\bf p}, {\bf t})}\,
z^{p_{\beta}+\delta_{\alpha \beta}-1}e^{\xi ({\bf t}_{\beta}, z)},
}
\\ \\
\displaystyle{\Psi_{\alpha \beta}^{\dag}({\bf p}, {\bf t};z)=
\epsilon_{\beta \alpha}({\bf p})\,
\frac{\tau_{\alpha \beta}({\bf p}, {\bf t}+[z^{-1}]_{\alpha})}{\tau ({\bf p}, {\bf t})}\,
z^{-p_{\alpha}+\delta_{\alpha \beta}-1}
e^{-\xi ({\bf t}_{\alpha}, z)}
}
\end{array}
\eeq
(here and below $\Psi^{\dag}$ does not mean the Hermitian conjugation).
In terms of the matrix Baker-Akhiezer functions, the bilinear identity (\ref{f7})
acquires the form
\beq\label{b2}
\oint \! dz \, \Psi ({\bf p}, {\bf t};z)\Psi^{\dag} ({\bf p}', {\bf t}';z)=0.
\eeq
Near $z=\infty$ the Baker-Akhiezer functions can be expanded into the series
\beq\label{b3}
\Psi_{\alpha \beta}({\bf p}, {\bf t};z)=\left (\delta_{\alpha \beta}+
\sum_{k\geq 1} w_{\alpha \beta}^{(k)}({\bf p}, {\bf t})z^{-k}\right )
z^{p_{\beta}} e^{\xi ({\bf t}_{\beta}, z)},
\eeq
\beq\label{b4}
\Psi_{\alpha \beta}^{\dag}({\bf p}, {\bf t};z)=\left (\delta_{\alpha \beta}+
\sum_{k\geq 1} v_{\alpha \beta}^{(k)}({\bf p}, {\bf t})z^{-k}\right )
z^{-p_{\alpha}} e^{-\xi ({\bf t}_{\alpha}, z)}. 
\eeq

It is proved in \cite{Teo11} that the Baker-Akhiezer function and its adjoint 
satisfy the auxiliary linear
equations
\beq\label{b5}
\begin{array}{l}
\,\,\,\,\, \p_{t_{\alpha , m}}\Psi ({\bf p}, {\bf t}; z)=B_{\alpha m} 
\Psi ({\bf p}, {\bf t}; z), 
\\ \\
-\p_{t_{\alpha , m}}\Psi^{\dag} ({\bf p}, {\bf t}; z)=
\Psi^{\dag} ({\bf p}, {\bf t}; z) B_{\alpha m} , 
\end{array}
\eeq
where $B_{\alpha m}$ is a matrix differential operator in 
$\displaystyle{\p_{t_1}\equiv \sum_{\alpha =1}^{N}\p_{t_{\alpha , 1}}}$.
In the second equation 
here it is assumed that the operators $\p_{t_1}$ entering $B_{\alpha m}$ act to the left
as $f\p_{t_1}=-\p_{t_1}f$.

\section{The matrix mKP hierarchy}

The matrix mKP hierarchy is obtained from the multicomponent one after the 
following restriction
of the time variables:
$$
t_{\alpha , m}=t_m \quad \mbox{for each $\alpha$ and $m$}, \quad
p_{\alpha}=p \quad \mbox{for each $\alpha$},
$$ 
so the evolution with respect to each $t_{\alpha , m}$ and $p_{\alpha}$
is the same and is defined by $t_m$, $p$ only. 
The corresponding vector fields are related as
$\p_{t_m}=\sum_{\alpha =1}^N \p_{t_{\alpha , m}}$,
$\p_p= \sum_{\alpha =1}^N \p_{p_{\alpha}}$.
In what follows we denote 
$$
\tau ({\bf p}, {\bf t}):= \tau ^p ({\bf t}), \quad
\epsilon_{\alpha \beta}({\bf p}):=\epsilon_{\alpha \beta}(p).
$$
The bilinear identity (\ref{f7}) acquires the form
\beq\label{f7a}
\sum_{\gamma =1}^N \epsilon_{\alpha \gamma}(p)\epsilon_{\beta \gamma}(p\! -\! n)
\! \oint \! dz z^{n+\delta_{\alpha \gamma}+\delta_{\beta \gamma}-2}
e^{\xi ({\bf t}_{\gamma}-{\bf t}_{\gamma}', z)} 
\tau_{\alpha \gamma}^{p}({\bf t}-[z^{-1}]_{\gamma})
\tau_{\gamma \beta}^{p-n}({\bf t}'+[z^{-1}]_{\gamma})=0
\eeq
with $n\geq 0$.
The Baker-Akhiezer functions (\ref{b1}) for the matrix mKP hierarchy are
\beq\label{m1}
\begin{array}{l}
\displaystyle{\Psi_{\alpha \beta}^{p}=
\epsilon_{\alpha \beta}(p)\,
\frac{\tau_{\alpha \beta}^p({\bf t}-[z^{-1}]_{\beta})}{\tau^p ({\bf t})}\,
z^{p+\delta_{\alpha \beta}-1}e^{\xi ({\bf t}, z)},
}
\\ \\
\displaystyle{\Psi_{\alpha \beta}^{\dag p}=
\epsilon_{\beta \alpha}(p)\,
\frac{\tau_{\alpha \beta}^p({\bf t}+[z^{-1}]_{\alpha})}{\tau^p ({\bf t})}\,
z^{-p+\delta_{\alpha \beta}-1}
e^{-\xi ({\bf t}, z)},
}
\end{array}
\eeq
where $\displaystyle{\xi ({\bf t}, z)=\sum_{k\geq 1}t_k z^k}$. Their expansions around
$z=\infty$ read
\beq\label{m2}
\Psi_{\alpha \beta}^{p}=\left (
\sum_{k\geq 0}w_{\alpha \beta}^{(k)}(p)z^{-k}\right )
z^{p+\delta_{\alpha \beta}-1}e^{\xi ({\bf t}, z)},
\eeq
\beq\label{m3}
\Psi_{\alpha \beta}^{\dag p}=\left (
\sum_{k\geq 0}v_{\alpha \beta}^{(k)}(p)z^{-k}\right )
z^{-p+\delta_{\alpha \beta}-1}e^{-\xi ({\bf t}, z)},
\eeq
where $w_{\alpha \beta}^{(0)}(p)=v_{\alpha \beta}^{(0)}(p)=\delta_{\alpha \beta}$
and we have suppressed the dependence on ${\bf t}$. 

In order to represent $w_{\alpha \beta}^{(k)}(p)$ and $v_{\alpha \beta}^{(k)}(p)$
explicitly for arbitrary $k$ we need some notation. Introduce the Schur polynomials
$h_k({\bf t})$ via the expansion
$$
\exp \Bigl (\sum_{k\geq 1}t_k z^k\Bigr )=\sum_{k\geq 0}h_k({\bf t})z^k
$$
(clearly, $h_0({\bf t})=1$ and $h_k({\bf t})=0$ for $k<0$). We also denote
$$
\begin{array}{l}
\tilde \p_{\alpha}=\Bigl \{\p_{t_{\alpha , 1}}, \frac{1}{2}\, \p_{t_{\alpha , 2}},
\frac{1}{3}\, \p_{t_{\alpha , 3}}, \ldots \Bigr \},
\end{array}
$$
so that $h_1(\tilde \p_{\alpha})=\p_{t_{\alpha , 1}}$, etc. Using the fact that
$\displaystyle{\tau ({\bf t}\pm [z^{-1}]_{\alpha})=
\exp \Bigl (\pm \sum_{k\geq 1}\frac{z^{-k}}{k}\, \p_{t_{\alpha , k}}
\Bigr )\tau ({\bf t})}$, we have
from (\ref{m1}):
\beq\label{m4}
w_{\alpha \beta}^{(k)}(p)=\frac{h_k(-\tilde \p_{\beta})\tau^p({\bf t})}{\tau^p({\bf t})}\,
\delta_{\alpha \beta}+\epsilon_{\alpha \beta}(p)
\frac{h_{k-1}(-\tilde \p_{\beta})\tau^p_{\alpha \beta}({\bf t})}{\tau^p({\bf t})}\,
(1-\delta_{\alpha \beta}),
\eeq
\beq\label{m4a}
v_{\alpha \beta}^{(k)}(p)=\frac{h_k(\tilde \p_{\alpha})\tau^p({\bf t})}{\tau^p({\bf t})}\,
\delta_{\alpha \beta}+\epsilon_{\beta \alpha}(p)
\frac{h_{k-1}(\tilde \p_{\alpha})\tau^p_{\alpha \beta}({\bf t})}{\tau^p({\bf t})}\,
(1-\delta_{\alpha \beta}).
\eeq
In particular, 
\beq\label{m5}
w^{(1)}_{\alpha \beta}(p)=\left \{
\begin{array}{l}
\displaystyle{\epsilon_{\alpha \beta}(p)
\, \frac{\tau_{\alpha \beta}^p({\bf t})}{\tau^p ({\bf t})}}\qquad
\!\!\! \mbox{if $\alpha \neq \beta$}
\\ \\
\displaystyle{-\, \frac{\p_{t_{\alpha , 1}}\tau^p ({\bf t})}{\tau^p ({\bf t})}} \qquad 
\mbox{if $\alpha = \beta$,}
\end{array}\right.
\eeq
$v_{\alpha \beta}^{(1)}(p)=-w_{\alpha \beta}^{(1)}(p)$ (the latter relation follows from
the bilinear identity in the form (\ref{b2}) at ${\bf t}={\bf t}'$, $p=p'$).

Let us introduce the matrix pseudo-difference wave operator
$$
W(p)=I+\sum_{k\geq 1}w^{(k)}(p)e^{-k\p_p},
$$
where $I$ is the unity $N\! \times \! N$ matrix and $w^{(k)}(p)$ are the same matrix
functions as in (\ref{m2}). In matrix elements we have
\beq\label{m6}
W_{\alpha \beta}(p)=\sum_{k\geq 0}w_{\alpha \beta}^{(k)}(p)e^{-k\p_p}.
\eeq
Clearly, the Baker-Akhiezer function $\Psi^p$ can be written as a result of 
action of the wave operator to the function $z^p$ times an exponential function:
\beq\label{m7}
\Psi^p= W(p)z^p \exp \Bigl (\sum_{\alpha =1}^N E_{\alpha}\xi ({\bf t}_{\alpha}, z)\Bigr ),
\eeq
where $E_{\alpha}$ is the $N\! \times \! N$ matrix with 1 on the diagonal $(\alpha , \alpha )$ 
component and zero elsewhere. 

We are going to show that the inverse wave operator $W^{-1}(p)$ is given by
\beq\label{m8}
W^{-1}(p)=\sum_{k\geq 0} e^{-k\p_p}\, v^{(k)}(p+1)
\eeq
with the matrices $v^{(k)}$ as in (\ref{m4a}). Indeed, we have:
$$
I=W(p)W^{-1}(p)=\sum_{k,k'\geq 0}w^{(k)}(p)e^{-(k+k')\p_p}v^{(k')}(p+1)
$$
$$
=\sum_{m\geq 0}\sum_{k=0}^m w^{(k)}(p)v^{(m-k)}(p\! +\! 1\! - \! m)e^{-m\p_p},
$$
so we need to show that
$$
\sum_{\gamma}\sum_{k=0}^mw_{\alpha \gamma}^{(k)}(p)
v_{\gamma \beta}^{(m-k)}(p\! +\! 1\! - \! m)=\delta_{m0}\delta_{\alpha \beta}.
$$
At $m=0$ this equality is obvious. At $m>0$, substituting (\ref{m4}), (\ref{m4a}), we get:
$$
\sum_{k=0}^m \epsilon_{\beta \alpha}(p\! +\! 1\! - \! m)h_k(-\tilde \p _{\alpha})
\tau^p({\bf t})h_{m-k-1}(\tilde \p_{\alpha})\tau_{\alpha \beta}^{p+1-m}({\bf t})
$$
$$
+\sum_{k=0}^m \epsilon_{\alpha \beta}(p)h_{k-1}(-\tilde \p _{\beta})\tau_{\alpha \beta}^p
({\bf t})h_{m-k}(\tilde \p_{\beta})\tau^{p+1-m}({\bf t})
$$
$$
+\sum_{\gamma \neq \alpha , \beta}\sum_{k=0}^m 
\epsilon_{\alpha \gamma}(p)\epsilon_{\beta \gamma}(p\! +\! 1\! - \! m)
h_{k-1}(-\tilde \p _{\gamma})\tau_{\alpha \gamma}^p ({\bf t})
h_{m-k-1}(\tilde \p_{\gamma})\tau_{\gamma \beta}^{p+1-m}({\bf t})\, =0
$$
for $\alpha \neq \beta$ and
$$
\sum_{k=0}^m h_k(-\tilde \p _{\alpha})\tau^p({\bf t})
h_{m-k}(\tilde \p_{\alpha})\tau^{p+1-m}({\bf t})
$$
$$
+\sum_{\gamma \neq \alpha}\sum_{k=0}^m
\epsilon_{\alpha \gamma}(p)\epsilon_{\alpha \gamma}(p\! +\! 1\! - \! m)
h_{k-1}(-\tilde \p _{\gamma})\tau_{\alpha \gamma}^p ({\bf t})
h_{m-k-1}(\tilde \p_{\gamma})\tau_{\gamma \alpha}^{p+1-m}({\bf t})\, =0
$$
for $\alpha =\beta$. 
One can check that these relations are precisely the ones that follow from 
the bilinear identity (\ref{f7a}) at $n=m-1$ and ${\bf t}={\bf t}'$, so the
formula (\ref{m8}) for the inverse wave operator is proved.  
Therefore, we can represent the adjoint Baker-Akhiezer function in the form
\beq\label{m9}
\Psi^{\dag p}=\exp \Bigl (-\sum_{\alpha}E_{\alpha}\xi ({\bf t}_{\alpha}, z)\Bigr )
z^{-p}W^{-1}(p-1),
\eeq
where the left action of the operators $e^{-\p_p}$ according to 
$fe^{-\p_p}=e^{\p_p}f$ is implied. 

In the next section we will show that the Baker-Akhiezer function $\Psi$ satisfies the 
auxiliary linear problem
\beq\label{m10}
\p_{t_1}\Psi = e^{\p_p}\Psi +u\Psi
\eeq
with some matrix function $u$. Therefore, since $B_{\alpha m}$ in the
auxiliary linear problems (\ref{b5}) are differential operators in $\p_{t_1}$,
$\p_{t_{\alpha , m}}\Psi$ can be expressed as a result of action of a difference
operator $A_{\alpha m}$ 
in $p$ that contains only non-negative powers of the shift operator $e^{\p_p}$.
This remark allows one to express $A_{\alpha m}$ in terms of the wave operator.
Indeed, differentiating equation (\ref{m7}), we have:
$$
\frac{\p \Psi^p}{\p t_{\alpha , m}}=\frac{\p W(p)}{\p t_{\alpha , m}}\, 
W^{-1}(p)\Psi^p +W(p)E_{\alpha}z^{m+p} \exp \Bigl (\sum_{\gamma}E_{\gamma}
\xi ({\bf t}_{\gamma}, z)\Bigr )
$$
$$
=\, \frac{\p W(p)}{\p t_{\alpha , m}}\,
W^{-1}(p)\Psi^p +W(p)E_{\alpha}e^{m\p_p}W^{-1}(p)\Psi^p.
$$
Therefore, 
$$
A_{\alpha m}(p)=\frac{\p W(p)}{\p t_{\alpha , m}}\,
W^{-1}(p)+W(p)E_{\alpha}e^{m\p_p}W^{-1}(p).
$$
Since the first term here contains strictly negative powers of $e^{\p_p}$,
one concludes that
\beq\label{m11}
\p_{t_{\alpha , m}}\Psi^p =A_{\alpha m}(p)\Psi^p, \qquad
A_{\alpha m}(p)=\Bigl (W(p)E_{\alpha}e^{m\p_p}W^{-1}(p)\Bigr )_+
\eeq
and
\beq\label{m12}
\p_{t_{\alpha , m}}W(p)=-\Bigl (W(p)E_{\alpha}e^{m\p_p}W^{-1}(p)\Bigr )_-
W(p),
\eeq
where $(\ldots )_{\pm}$ denotes the part of a pseudo-difference operator containing 
only non-\-ne\-ga\-ti\-ve (respectively, negative) powers of $e^{\p_p}$. In a 
similar way, differentiating (\ref{m9}), we obtain the auxiliary linear problems
for the adjoint Baker-Akhiezer function:
\beq\label{m13}
-\p_{t_{\alpha , m}}\Psi^{\dag p}=\Psi^{\dag p}A_{\alpha m}(p-1).
\eeq

One can also introduce the Lax operator
\beq\label{m14}
L(p)=W(p)e^{\p_p}W^{-1}(p),
\eeq
with $\Psi$ being its eigenfunction:
\beq\label{m15}
L(p)\Psi^p=z\Psi^p.
\eeq
The compatibility of (\ref{m11}) and (\ref{m15}) implies the Lax equation
\beq\label{m16}
\p_{t_{\alpha , m}}L(p)=[A_{\alpha m}(p), L(p)].
\eeq

We have thus obtained the auxiliary linear problems and Lax equations
for the multicomponent hierarchy. They simplify for the matrix hierarchy:
\beq\label{m17}
\begin{array}{l}
\,\, \, \p_{t_m}\Psi^p =A_m(p)\Psi^p,
\\ \\
-\p_{t_m}\Psi^{\dag p}=\Psi^{\dag p}A_m(p-1),
\end{array}
\eeq
where
\beq\label{m18}
A_m(p)=\Bigl (W(p)e^{m\p_p}W^{-1}(p)\Bigr )_+
\eeq
(we use the fact that $\displaystyle{\sum_{\alpha =1}^{N}E_{\alpha}=I}$).
In particular, for $m=1$ we have:
\beq\label{m19}
\begin{array}{l}
\,\, \, \p_{t_1}\Psi^p =\Psi^{p+1}+\Bigl (w^{(1)}(p)-w^{(1)}(p+1)\Bigr )\Psi^p ,
\\ \\
-\p_{t_1}\Psi^{\dag p}=\Psi^{\dag p-1}+\Psi^{\dag p}\Bigl (w^{(1)}(p-1)-w^{(1)}(p)\Bigr ).
\end{array}
\eeq
In fact the very derivation of (\ref{m11}) and (\ref{m17}), (\ref{m18}) 
was based on equations (\ref{m19}) (see (\ref{m10})).
In the next section we will derive these equations in an independent way as 
a consequence of the bilinear identity. It should be noted that in \cite{KZ95}
the linear problems (\ref{m19}) for Baker-Akhiezer functions on Riemann 
surfaces were obtained using algebro-geometric reasoning.

\section{Auxiliary linear problem for derivative with respect to $t_1$}

In this section we show that the auxiliary linear problems (\ref{m19})
are in fact equivalent to a consequence of the bilinear identity (\ref{f7}).
We start from the linear problem for $\Psi$ writing it in components in the form
\beq\label{a1}
\Psi_{\alpha \beta}^{p+1}=\p_{t_1}\Psi_{\alpha \beta}^p +\sum_{\gamma}
\Bigl (w_{\alpha \gamma}^{(1)}(p+1)-w_{\alpha \gamma}^{(1)}(p)\Bigr )
\Psi_{\gamma \beta}^p.
\eeq
We recall that $w_{\alpha \gamma}^{(1)}(p)$ is given by (\ref{m5}). 
Consider first the case $\alpha \neq \beta$. Substituting (\ref{m1}) for
$\Psi$, we write (\ref{a1}) in the form
$$
\epsilon_{\alpha \beta}(p+1)
\frac{\tau_{\alpha \beta}^{p+1}({\bf t}-[z^{-1}]_{\beta})}{\tau^{p+1}({\bf t})}=
\epsilon_{\alpha \beta}(p)
\frac{\tau_{\alpha \beta}^{p}({\bf t}-[z^{-1}]_{\beta})}{\tau^{p}({\bf t})}+
\epsilon_{\alpha \beta}(p)\p_{t_1}\! \left 
(\frac{\tau_{\alpha \beta}^{p}({\bf t}-[z^{-1}]_{\beta}}{\tau^{p}({\bf t})}\right )\! z^{-1}
$$
$$
+\sum_{\gamma \neq \alpha}\epsilon_{\alpha \gamma}(p+1)\epsilon_{\gamma \beta}(p)
\frac{\tau_{\alpha \gamma}^{p+1}({\bf t})
\tau_{\gamma \beta}^{p}({\bf t}-[z^{-1}]_{\beta})}{\tau^{p+1}({\bf t})\tau^p({\bf t})}
\, z^{\delta_{\gamma \beta}-1}
$$
$$
-\sum_{\gamma \neq \alpha}\epsilon_{\alpha \gamma}(p)\epsilon_{\gamma \beta}(p)
\frac{\tau_{\alpha \gamma}^{p}({\bf t})
\tau_{\gamma \beta}^{p}({\bf t}-[z^{-1}]_{\beta})}{(\tau^{p}({\bf t}))^2}
\, z^{\delta_{\gamma \beta}-1}
$$
$$
-\epsilon_{\alpha \beta}(p)\left (
\frac{\p_{t_{\alpha , 1}}\tau^{p+1}({\bf t})}{\tau^{p+1}({\bf t})}-
\frac{\p_{t_{\alpha , 1}}\tau^{p}({\bf t})}{\tau^{p}({\bf t})}\right )
\frac{\tau_{\alpha \beta}^p ({\bf t}-[z^{-1}]_{\beta})}{\tau^p({\bf t})}\, z^{-1}.
$$
After some obvious transformations, separating the terms with the denominator
$(\tau^p({\bf t}))^2$, we rewrite this as
$$
-\epsilon_{\alpha \beta}(p+1)
\frac{\tau_{\alpha \beta}^{p+1}({\bf t}-[z^{-1}]_{\beta})}{\tau^{p+1}({\bf t})}+
\epsilon_{\alpha \beta}(p)
\frac{\tau_{\alpha \beta}^{p}({\bf t}-[z^{-1}]_{\beta})}{\tau^{p}({\bf t})}+
\epsilon_{\alpha \beta}(p)
\frac{\p_{t_1}\tau_{\alpha \beta}^p({\bf t}-[z^{-1}]_{\beta})}{\tau^p({\bf t})}\, z^{-1}
$$
$$
+\epsilon_{\alpha \beta}(p+1)\frac{\tau_{\alpha \beta}^{p+1}({\bf t})
\tau^p({\bf t}-[z^{-1}]_{\beta})}{\tau^{p+1}({\bf t}) \tau^p({\bf t})}
-\epsilon_{\alpha \beta}(p)\frac{\p_{t_{\alpha , 1}}\tau^{p+1}({\bf t})\,
\tau_{\alpha \beta}^p({\bf t}-[z^{-1}]_{\beta})}{\tau^{p+1}({\bf t}) \tau^p({\bf t})}\,
z^{-1}
$$
$$
+z^{-1}\sum_{\gamma \neq \alpha , \beta}
\epsilon_{\alpha \gamma}(p+1)\epsilon_{\gamma \beta}(p)
\frac{\tau_{\alpha \gamma}^{p+1}({\bf t})
\tau_{\gamma \beta}^{p}({\bf t}-[z^{-1}]_{\beta})}{\tau^{p+1}({\bf t})\tau^p({\bf t})}
$$
$$
+\frac{\epsilon_{\alpha \beta}(p)}{(\tau^p ({\bf t}))^2}\left \{
-\tau_{\alpha \beta}^p({\bf t}-[z^{-1}]_{\beta})\p_{t_1}\tau^p({\bf t})z^{-1}
-\tau_{\alpha \beta}^p ({\bf t})\tau^p({\bf t}-[z^{-1}]_{\beta})
\phantom{\sum_{\gamma \neq \alpha }^N}
\right.
$$
$$
\left.  + z^{-1}\p_{t_{\alpha , 1}}\tau^p({\bf t})\,
\tau_{\alpha \beta}^{p}({\bf t}-[z^{-1}]_{\beta})-
z^{-1}\!\!\sum_{\gamma \neq \alpha , \beta}
\frac{\epsilon_{\alpha \gamma}(p)\epsilon_{\gamma \beta}(p)}{\epsilon_{\alpha \beta}(p)}\,
\tau_{\alpha \gamma}^p({\bf t})\tau_{\gamma \beta}^p({\bf t}-[z^{-1}]_{\beta})\right \}=0.
$$
Let us transform the expression in the brackets $\{\ldots \}$:
$$
\left \{\phantom{\int}\!\!\!\! \ldots \phantom{\int}\!\!\!\! \right \}=
-z^{-1} 
\tau_{\alpha \beta}^p({\bf t}-[z^{-1}]_{\beta})\p_{t_{\beta , 1}}\tau^p({\bf t})-
\tau_{\alpha \beta}^p ({\bf t})\tau^p({\bf t}-[z^{-1}]_{\beta})
$$
$$
-z^{-1}\!\! \sum_{\gamma \neq \alpha , \beta}\left [
\tau_{\alpha \beta}^p ({\bf t}-[z^{-1}]_{\beta})\p_{t_{\gamma , 1}}\tau^p({\bf t})+
\frac{\epsilon_{\alpha \gamma}(p)\epsilon_{\gamma \beta}(p)}{\epsilon_{\alpha \beta}(p)}\,
\tau_{\alpha \gamma}^p({\bf t})\tau_{\gamma \beta}^p({\bf t}-[z^{-1}]_{\beta})
\right ].
$$
In the first line, we use the Hirota equation (\ref{f9}) with $\mu =\infty$,
$\nu =z$.
In the second line, we use the Hirota equation (\ref{f8}) with $\mu =z$. The result is
$$
\left \{\phantom{\int}\!\!\!\! \ldots \phantom{\int}\!\!\!\! \right \}=
-\tau^p({\bf t})\left [z^{-1}\! \sum_{\gamma \neq \alpha}\p_{t_{\gamma , 1}}
\tau_{\alpha \beta}^p({\bf t}-[z^{-1}]_{\beta})+
\tau_{\alpha \beta}^p({\bf t}-[z^{-1}]_{\beta})\right ].
$$
Substituting this back, we obtain, after some cancellations:
$$
-\epsilon_{\alpha \beta}(p+\! 1)
\frac{\tau_{\alpha \beta}^{p+1}({\bf t}-[z^{-1}]_{\beta})}{\tau^{p+1}({\bf t})}+
\epsilon_{\alpha \beta}(p)
\frac{\p_{t_{\alpha , 1}}\tau_{\alpha \beta}^p
({\bf t}-[z^{-1}]_{\beta})}{\tau^p({\bf t})}\, z^{-1}
$$
$$
+\epsilon_{\alpha \beta}(p+\! 1)\frac{\tau_{\alpha \beta}^{p+1}({\bf t})
\tau^p({\bf t}-[z^{-1}]_{\beta})}{\tau^{p+1}({\bf t}) \tau^p({\bf t})}
-\epsilon_{\alpha \beta}(p)\frac{\p_{t_{\alpha , 1}}\tau^{p+1}({\bf t})\,
\tau_{\alpha \beta}^p({\bf t}-[z^{-1}]_{\beta})}{\tau^{p+1}({\bf t}) \tau^p({\bf t})}\,
z^{-1}
$$
$$
+z^{-1}\!\!\sum_{\gamma \neq \alpha , \beta}
\epsilon_{\alpha \gamma}(p+1)\epsilon_{\gamma \beta}(p)
\frac{\tau_{\alpha \gamma}^{p+1}({\bf t})
\tau_{\gamma \beta}^{p}({\bf t}-[z^{-1}]_{\beta})}{\tau^{p+1}({\bf t})\tau^p({\bf t})}=0
$$
or
$$
z\epsilon_{\alpha \beta}(p+\! 1)\tau_{\alpha \beta}^{p+1}({\bf t})
\tau^p({\bf t}-[z^{-1}]_{\beta})-
z\epsilon_{\alpha \beta}(p+\! 1)\tau^p({\bf t})
\tau^{p+1}_{\alpha \beta}({\bf t}-[z^{-1}]_{\beta})
$$
$$
+\, \epsilon_{\alpha \beta}(p)\p_{t_{\alpha , 1}}
\tau^{p}_{\alpha \beta}({\bf t}-[z^{-1}]_{\beta})\tau^{p+1}({\bf t})-
\epsilon_{\alpha \beta}(p)\p_{t_{\alpha , 1}}\tau^{p+1}({\bf t})
\tau^{p}_{\alpha \beta}({\bf t}-[z^{-1}]_{\beta})
$$
$$
+\sum_{\gamma \neq \alpha , \beta}
\epsilon_{\alpha \gamma}(p+1)\epsilon_{\gamma \beta}(p)
\tau_{\alpha \gamma}^{p+1}({\bf t})
\tau_{\gamma \beta}^{p}({\bf t}-[z^{-1}]_{\beta})=0.
$$
One can see that this relation is equivalent to the bilinear identity
(\ref{f7a}) taken at $n=1$, $\alpha \neq \beta$, ${\bf t}'={\bf t}-[\mu^{-1}]_{\beta}$ 
(with $\mu =z$ in the end). 

We now pass to the case $\alpha = \beta$ in (\ref{a1}):
$$
z\, \frac{\tau^{p+1}({\bf t}-[z^{-1}]_{\alpha})}{\tau^{p+1}({\bf t})}=
z\,
\frac{\tau^{p}({\bf t}-[z^{-1}]_{\alpha})}{\tau^{p}({\bf t})}+
\p_{t_1}\! \left 
(\frac{\tau^{p}({\bf t}-[z^{-1}]_{\alpha}}{\tau^{p}({\bf t})}\right )
$$
$$
+z^{-1}\! \sum_{\gamma \neq \alpha}\epsilon_{\alpha \gamma}(p+1)\epsilon_{\gamma \alpha}(p)
\frac{\tau_{\alpha \gamma}^{p+1}({\bf t})
\tau_{\gamma \alpha}^{p}({\bf t}-[z^{-1}]_{\alpha})}{\tau^{p+1}({\bf t})\tau^p({\bf t})}
+z^{-1}\!\sum_{\gamma \neq \alpha}
\frac{\tau_{\alpha \gamma}^{p}({\bf t})
\tau_{\gamma \alpha}^{p}({\bf t}-[z^{-1}]_{\alpha})}{(\tau^{p}({\bf t}))^2}
$$
$$
-\left (
\frac{\p_{t_{\alpha , 1}}\tau^{p+1}({\bf t})}{\tau^{p+1}({\bf t})}-
\frac{\p_{t_{\alpha , 1}}\tau^{p}({\bf t})}{\tau^{p}({\bf t})}\right )
\frac{\tau^p ({\bf t}-[z^{-1}]_{\alpha})}{\tau^p({\bf t})}.
$$
Separating the terms with the denominator
$(\tau^p({\bf t}))^2$, we rewrite this as
$$
-z\, \frac{\tau^{p+1}({\bf t}-[z^{-1}]_{\alpha})}{\tau^{p+1}({\bf t})}+
z\,
\frac{\tau^{p}({\bf t}-[z^{-1}]_{\alpha})}{\tau^{p}({\bf t})}\, +
\frac{\p_{t_1}\tau^p({\bf t}-[z^{-1}]_{\alpha})}{\tau^{p}({\bf t})}
$$
$$
+z^{-1}\! \sum_{\gamma \neq \alpha}\epsilon_{\alpha \gamma}(p+1)\epsilon_{\gamma \alpha}(p)
\frac{\tau_{\alpha \gamma}^{p+1}({\bf t})
\tau_{\gamma \alpha}^{p}({\bf t}-[z^{-1}]_{\alpha})}{\tau^{p+1}({\bf t})\tau^p({\bf t})}
-\frac{\p_{t_{\alpha , 1}}\tau^{p+1}({\bf t})
\tau^p({\bf t}-[z^{-1}]_{\alpha})}{\tau^{p+1}({\bf t})\tau^p({\bf t})}
$$
$$
+\frac{1}{(\tau^p({\bf t}))^2}
\left \{\sum_{\gamma \neq \alpha}\Bigl (
z^{-1}\tau_{\alpha \gamma}^p({\bf t})\tau_{\gamma \alpha}^p({\bf t}-[z^{-1}]_{\alpha})
-\tau^p({\bf t}-[z^{-1}]_{\alpha})\p_{t_{\gamma , 1}}\tau^p({\bf t})
\Bigr )
\right \}=0.
$$
Using the Hirota equation (\ref{f11}), we obtain:
$$
\left \{\phantom{\int}\!\!\!\! \ldots \phantom{\int}\!\!\!\! \right \}=
-\tau^p({\bf t})\sum_{\gamma \neq \alpha}\p_{t_{\gamma , 1}}
\tau^p({\bf t}-[z^{-1}]_{\alpha}),
$$
so the previous expression acquires the form
$$
-z\, \frac{\tau^{p+1}({\bf t}-[z^{-1}]_{\alpha})}{\tau^{p+1}({\bf t})}+
z\,
\frac{\tau^{p}({\bf t}-[z^{-1}]_{\alpha})}{\tau^{p}({\bf t})}\, +
\frac{\p_{t_{\alpha , 1}}\tau^p({\bf t}-[z^{-1}]_{\alpha})}{\tau^{p}({\bf t})}
$$
$$
-\, \frac{\p_{t_{\alpha , 1}}\tau^{p+1}({\bf t})
\tau^p({\bf t}-[z^{-1}]_{\alpha})}{\tau^{p+1}({\bf t})\tau^p({\bf t})}
+z^{-1}\! \sum_{\gamma \neq \alpha}\epsilon_{\alpha \gamma}(p+1)\epsilon_{\gamma \alpha}(p)
\frac{\tau_{\alpha \gamma}^{p+1}({\bf t})
\tau_{\gamma \alpha}^{p}({\bf t}-[z^{-1}]_{\alpha})}{\tau^{p+1}({\bf t})\tau^p({\bf t})}=0
$$
or
$$
z^2\tau^{p+1}({\bf t})\tau^{p}({\bf t}-[z^{-1}]_{\alpha})-z^2
\tau^{p+1}({\bf t}-[z^{-1}]_{\alpha})\tau^{p}({\bf t})
$$
$$
+z\tau^{p+1}({\bf t})\p_{t_{\alpha , 1}}\tau^{p}({\bf t}-[z^{-1}]_{\alpha})-
z\p_{t_{\alpha , 1}}\tau^{p+1}({\bf t})\tau^{p}({\bf t}-[z^{-1}]_{\alpha})
$$
$$
+\sum_{\gamma \neq \alpha}\epsilon_{\alpha \gamma}(p+1)\epsilon_{\gamma \alpha}(p)
\tau_{\alpha \gamma}^{p+1}({\bf t})
\tau_{\gamma \alpha}^{p}({\bf t}-[z^{-1}]_{\alpha})=0.
$$
One can see that this relation is equivalent to the bilinear identity
(\ref{f7a}) taken at $n=1$, $\alpha =\beta$, ${\bf t}'={\bf t}-[\mu^{-1}]_{\alpha}$ 
(with $\mu =z$ in the end). 

The second equation in (\ref{m19}), which we write in components in the form
\beq\label{a2}
\Psi^{\dag p}_{\alpha \beta}=-\p_{t_1}\Psi^{\dag p+1}_{\alpha \beta}+
\sum_{\gamma}\Psi^{\dag p+1}_{\alpha \gamma}\Bigl (w_{\gamma \beta}^{(1)}(p+1)-
w_{\gamma \beta}^{(1)}(p)\Bigr )
\eeq
can be processed in a similar way, using the Hirota equations (\ref{f8}),
(\ref{f10}), (\ref{f11}). 

Finally, we would like to remark that one can interpret the linear problems (\ref{m19}) 
also in a different way. Namely, the equation $\p_{t_1}\Psi =e^{\p_p}\Psi +u\Psi$
can be read as $\Psi^{p+1}=\p_{t_1}\Psi^p -u\Psi^p$, and in this form it can be 
regarded as describing an extension of the matrix KP hierarchy to the discrete flow $p$
generated by the first order differential operator $\p_{t_1}-u$.

\section{Conclusion}

In this paper we have studied the matrix and multicomponent mKP hierarchies
from the point of view of fermionic formalism and bilinear identities. 
The pseudo-difference wave operator has been introduced and the auxiliary linear
problems for the Baker-Akhiezer function and its adjoint have been derived.
The Lax representation of the hierarchy in terms of pseudo-difference operators 
has been obtained.

It should be noted that even more general matrix hierarchy than mKP exists.
It is the matrix or non-abelian 
(and multicomponent) Toda hierarchy, in which in addition to 
$t_{\alpha , m}$ with $m>0$ there is yet another 
infinite set of continuous times $t_{\alpha , m}$ with $m<0$. It would be interesting to
extend the approach developed in this paper to this more general case. 
The algebro-geometric solutions to the matrix Toda hierarchy were discussed in
\cite{KZ95,K81}.

\section*{Acknowledgments}

This work was funded by the Russian Academic Excellence
Project `5-100'. This work was supported in part by RFBR grant
18-01-00461.

\end{document}